\documentclass[12pt,prb]{revtex4}

\usepackage{setspace}
\usepackage{graphicx}
\usepackage{dcolumn}
\usepackage{bm}

\newcommand{\mrd}{\hspace{0.6mm}}
\begin{document}

\preprint{}

\title{Symmetry of $k\cdot p$ Hamiltonian in pyramidal InAs/GaAs quantum dots: Application
to the calculation of electronic structure}

 \author{Nenad Vukmirovi\'c}
 \email{eennv@leeds.ac.uk}
  \author{Dragan Indjin}
 \author{Vladimir D. Jovanovi\'c}
 \author{Zoran Ikoni\'c}
 \author{Paul Harrison}
 \affiliation{
 School of Electronic and Electrical Engineering,
 University of Leeds, Leeds LS2 9JT, United Kingdom
 }

 
\singlespacing

\date{\today}

\begin{abstract}
A method for the calculation of the electronic structure of pyramidal self-assembled 
InAs/GaAs quantum dots is presented. The method is based on exploiting the  
$\overline{C}_{4}$ symmetry of the 8-band $k\cdot p$ Hamiltonian with the strain 
taken into account via the continuum mechanical model. 
The operators representing symmetry group elements were represented in the plane wave basis
and the group projectors were used to find the symmetry adapted basis in which the corresponding
Hamiltonian matrix is block diagonal with four blocks of approximately equal size. 
The quantum number of total quasi-angular momentum is introduced and the states are 
classified according to its value. Selection rules for interaction with electromagnetic field in the
dipole approximation are derived. 
The method was applied to calculate electron and hole quasibound states in a periodic array of 
vertically stacked pyramidal self-assembled InAs/GaAs quantum dots for different values of
the distance between the dots and external axial magnetic field. As the distance between the
dots in an array is varied, an interesting effect of simultaneous change of ground hole state symmetry, 
type and the sign of miniband effective mass is predicted. This effect is explained in terms of the change 
of biaxial strain. It is also found that the magnetic field splitting of Kramer's double 
degenerate states is most prominent for the first and second excited state in the conduction band and that
the magnetic field can both separate otherwise overlapping minibands and concatenate otherwise nonoverlapping
minibands.
\end{abstract}

\pacs{73.21.La}
\maketitle

\section*{Introduction}
Semiconductor quantum dots made by Stranski-Krastanow growth have attracted 
great interest over the past few years from the view of fundamental physics, as well as
due to their application in
optoelectronic and microelectronic devices. In order to understand the physics
of quantum dots and model and design
such devices the electronic structure needs to be accurately known. The large
lattice mismatch between InAs and GaAs has enabled the fabrication of quantum
dots putting it at the forefront of both theoretical
and experimental research. Different quantum dot shapes (such as pyramid\cite{apl65-1421}, lens\cite{apl63-3203} 
 and disk\cite{sci291-451}) of InAs/GaAs self-assembled quantum dots are often reported. 

A range of theoretical 
approaches has been used so far to calculate the energy levels in self-assembled
quantum dots - effective mass \cite{ssc92-437,prb61-10959, prb61-13840, prb54-R2300}, $k\cdot p$ 
\cite{prb59-5688, prl80-3579, prb56-4696, prb57-7190} and the pseudopotential
method \cite{prb59-5678, prb71-045318}. In quantum dots with cylindrical symmetry, symmetry considerations have been 
applied to effectively reduce the geometry of the problem from three-dimensional to two-dimensional, both in the effective mass and the $k\cdot p$ method (within the axial approximation) \cite{prb65-165333}. The possible symmetries of the states in hexagonal III-nitride quantum dots have recently
been determined \cite{pssb241-2938}.
The symmetry of the pyramid has been used in the effective mass calculation \cite{prb68-235308} to reduce
the size of the corresponding Hamiltonian matrix, however in none of the $k\cdot p$ 
calculations of pyramidal quantum dots has the explicit use of the symmetry of the Hamiltonian been reported.
The aim of this paper is to exploit the symmetry in $k\cdot p$
calculation of the electronic structure of pyramidal InAs/GaAs quantum dots.

The symmetry of a pyramidal InAs/GaAs quantum dot when a full atomistic structure is considered is $\overline{C}_{2v}$ and is lower than the symmetry of the dot's geometrical shape\cite{prb71-045318}. Due to the computational complexity of the pseudopotential methods that take into account the atomistic nature of the structure, one often employs the $k\cdot p$ method which is considered to be a reliable tool for modeling the electronic structure of quantum dots despite its known limitations\cite{prb59-5688}. The symmetry of the $k\cdot p$ model itself is the symmetry group of the zincblende crystal lattice. When the model is applied to pyramidal quantum dots the symmetry group is the intersection of the geometrical symmetry of the pyramid shape and the zincblende bulk symmetry. Since the pyramid shape symmetry group is a subgroup of the symmetry of the zincblende crystal lattice, it follows that
the symmetry group of the model is the double $\overline{C}_4$ group \cite{Elliott}. Two different approaches are used to calculate the strain distribution in quantum dots - the continuum mechanical \cite{prb57-7190, prb59-5688} and the valence force field model \cite{prb56-4696, prb59-5688}. When the strain distribution is incorporated in the $k\cdot p$ method, the continuum mechanical model preserves the $\overline{C}_4$ symmetry, while the valence force field model, due to its atomistic nature, breaks it \cite{prb56-4696, prb59-5688}. Nevertheless, the comparisons of the two models have shown that they give similar results \cite{prb59-5688, jap92-5819}. In this paper, we apply the 8-band $k\cdot p$ method with the strain taken into account via the continuum mechanical model. We shall refer to this in the rest of the text for brevity as the model. Therefore, the symmetry group of this model is $\overline{C}_4$. All the results presented in Sec.~\ref{sec:results} are strictly valid in the framework of such an idealized model.

In previous years, arrays of vertically stacked quantum dots with 10 or more layers have been reported \cite{prl76-952, prb60-16680}. 
The quantum wire behavior of such structures was theoretically investigated in Ref.~\onlinecite{prl80-3579}.
Most of the theoretical investigations of the InAs quantum dots in a magnetic field have focused on the dots with a parabolic confinement potential
\cite{prb58-3561, prb55-4580, prb55-9275, prb69-201308(R), prb68-115310, prb69-161308(R), prb70-081314(R)} or just on the conduction band states \cite{prb63-195311, prb68-235308}. More recently, the influence of a magnetic field on pyramidal single quantum dots has been investigated in the framework of the $k\cdot p$ method \cite{prb70-235337}. 

This paper is organized as follows. In Sec.~\ref{sec:pwm} the relations necessary to apply the plane wave
expansion method\cite{prb62-15851} to 8-band $k\cdot p$ calculation of quantum dots based on materials with zincblende crystal
symmetry in the presence of strain and external axial magnetic field are given.
In Sec.~\ref{sec:symmetry} the representations of the symmetry group
elements, the reduction of the representation to its irreducible constituents and finally the symmetry adapted basis
are found. The states are classified according to their symmetry and selection rules for absorption of electromagnetic 
radiation in the dipole approximation are found. It is also shown how symmetry considerations can be incorporated
into the plane wave method. Finally, in Sec.~\ref{sec:results} the method presented is applied to calculate the electron and hole quasibound states in a 
periodic array of vertically stacked pyramidal self-assembled InAs/GaAs quantum dots
for different values of the period of the structure and in the presence of an external axial magnetic field.

\section {The plane wave method}\label{sec:pwm}
In the presence of an axial magnetic field the total 8-band $k\cdot p$ Hamiltonian is a sum of 
the kinetic part of the Hamiltonian $\hat{H}_k$, the strain part $\hat{H}_s$, the modification 
of the kinetic part due to magnetic field $\hat{H}_B$ and the Zeeman part $\hat{H}_Z$
\begin{equation}
\hat{H}=\hat{H}_k+\hat{H}_s+\hat{H}_B+\hat{H}_Z.
\end{equation}
The state of the system within the framework of the 8-band $k\cdot p$ method is given as a sum of slowly varying 
envelope functions $\psi_i({\bm r})$ multiplied by the bulk Bloch functions $|i\rangle$ (Eq.~\ref{eq:basis}), i.e.
\begin{equation}\label{eq:state}
|\Psi\rangle=\sum_{i=1}^8 \psi_i({\bm r})|i\rangle.
\end{equation} 
The eigenvalue problem of the 8-band $k\cdot p$ Hamiltonian $\hat{H}$ can therefore be written as
\begin{equation}\label{eq:problem1}
\sum_{j=1}^{8}H_{ij}\psi_j({\bm r})=E\psi_i({\bm r}).
\end{equation}
The basis of Bloch functions $|J,J_z\rangle$ (where $\hat{\bm{J}}$ is the total angular momentum of the Bloch function)
used to represent the Hamiltonian is given by \cite{prb41-11992}
\begin{eqnarray}\label{eq:basis}
|1\rangle&=&|\frac{1}{2},-\frac{1}{2}\rangle=|S\downarrow\rangle,\nonumber\\ \nonumber
|2\rangle&=&|\frac{1}{2},\frac{1}{2}\rangle=|S\uparrow\rangle,\\ \nonumber
|3\rangle&=&|\frac{3}{2},-\frac{3}{2}\rangle=-\frac{i}{\sqrt{2}}|(X-iY)\downarrow\rangle,\\ \nonumber
|4\rangle&=&|\frac{3}{2},-\frac{1}{2}\rangle=\frac{i}{\sqrt{6}}|(X-iY)\uparrow\rangle+i\sqrt{\frac{2}{3}}|Z\downarrow\rangle,\\ \nonumber
|5\rangle&=&|\frac{3}{2},\frac{1}{2}\rangle=-\frac{i}{\sqrt{6}}|(X+iY)\downarrow\rangle+i\sqrt{\frac{2}{3}}|Z\uparrow\rangle,\\ \nonumber
|6\rangle&=&|\frac{3}{2},\frac{3}{2}\rangle=\frac{i}{\sqrt{2}}|(X+iY)\uparrow\rangle,\\ \nonumber
|7\rangle&=&|\frac{1}{2},-\frac{1}{2}\rangle=-\frac{i}{\sqrt{3}}|(X-iY)\uparrow\rangle+\frac{i}{\sqrt{3}}|Z\downarrow\rangle,\\ 
|8\rangle&=&|\frac{1}{2},\frac{1}{2}\rangle=-\frac{i}{\sqrt{3}}|(X+iY)\downarrow\rangle-\frac{i}{\sqrt{3}}|Z\uparrow\rangle.
\end{eqnarray}
The kinetic and the strain part of the Hamiltonian in the basis (\ref{eq:basis}) are given in Ref.~\onlinecite{prb41-11992} and
the matrix elements of the 8-band $k\cdot p$ Hamiltonian
in the case of a magnetic field of arbitrary direction are given in Ref.~\onlinecite{prb38-6151}. In the case of the axial magnetic 
field that we are interested in, the influence of the magnetic field $B$ can be taken into account by replacing $k_i$ with
$k_i+\frac{e}{\hbar}A_i$ in the kinetic part of the Hamiltonian, where $A_i$ is the $i-$th component of the magnetic 
vector potential, and by adding the Zeeman term $\hat{H}_Z$ (Ref.~\onlinecite{prb38-6151}). We take the symmetric gauge for 
the vector potential ${\bm A}=\frac{1}{2}B\left(-y,x,0\right)$.



\begin{figure}[htbp]
\vspace{1cm}
\includegraphics[width=0.5\textwidth]{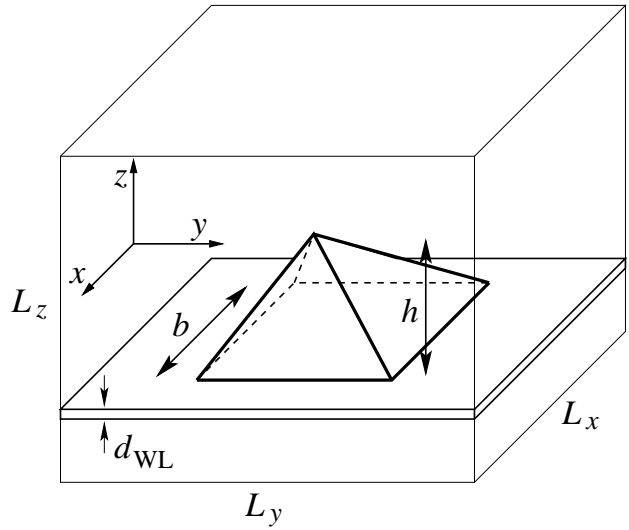}
\caption[]{Quantum dot geometry. The width of the base is $b$, the height $h$, the wetting layer width $d_{WL}$.
 The embedding box sides are $L_x$, $L_y$ and $L_z$.
The center of the pyramid base is taken as the origin of the coordinate system.}
\label{fig:1QD}
\end{figure}

The plane wave method is based on embedding the quantum dot in a box of sides $L_x$, $L_y$ and $L_z$ (Fig.~\ref{fig:1QD})
and assuming the envelope functions are a linear combination of plane waves
\begin{equation}\label{eq:problem2}
\psi_i({\bm r}) = \sum_{{\bm k}} A_{i,{\bm k}}\exp\left(i{\bm k}\cdot{\bm r}\right),
\end{equation}
with the coefficients $A_{i,{\bm k}}$ to be determined. The wave vectors taken in a summation are
given by ${\bm k}=2\pi\left(\frac{m_x}{L_x},\frac{m_y}{L_y},\frac{m_z}{L_z}\right)$ ($m_x\in\{-n_x,\ldots,n_x\}$,
 $m_y\in\{-n_y,\ldots,n_y\}$, $m_z\in\{-n_z,\ldots,n_z\}$).
The number of plane waves taken is thus $N=8(2n_x+1)(2n_y+1)(2n_z+1),$ 
where $2n_t+1$ is the number of plane waves per dimension $t$ ($t\in\{x,y,z\}$). 
Due to the symmetry of the pyramid, the embedding box sides $L_x$ and $L_y$ are taken to be equal ($L_x=L_y$),
as well as the number of plane waves per dimensions $x$ and $y$ ($n_x=n_y$). 

After multiplying equation (\ref{eq:problem1}) from the left by 
$\frac{1}{(2\pi)^3}\int_V{\rm d}^3{\bm r}\exp\left(-i{\bm q}\cdot{\bm r}\right),$
where the integration goes over the volume of the embedding box, using 
(\ref{eq:problem2}) and the identity
\begin{equation}
\int_V{\rm d}^3{\bm r}\exp\left[2i\pi\left(\frac{m_xx}{L_x}+\frac{m_yy}{L_y}+\frac{m_zz}{L_z}\right)\right]=
L_xL_yL_z\delta_{m_x,0}\delta_{m_y,0}\delta_{m_z,0},
\end{equation}
one arrives at 
\begin{equation}\label{eq:eigen}
\sum_{j,{\bm k}}h_{ij}({\bm q},{\bm k})A_{j,{\bm k}}=EA_{i,{\bm q}},
\end{equation}
where 
\begin{equation}\label{eq:fur}
h_{ij}({\bm q},{\bm k})=\frac{1}{V}\int_V{\rm d}^3{\bm r}\exp\left(-i{\bm q}\cdot{\bm r}\right)H_{ij}\exp\left(i{\bm k}\cdot{\bm r}\right).
\end{equation}
The material parameters in $\hat{H}_k$, $\hat{H}_s$, $\hat{H}_B$ and $\hat{H}_Z$ are all spatially varying functions since
they have different values in the quantum dot and the matrix. Therefore the Hamiltonian matrix 
$h$ can loose hermiticity unless the proper recipe for the order of differential and multiplication
operators is chosen. This recipe is given by
\begin{eqnarray}
f({\bm r})\partial_i\partial_j&\to&\frac{1}{2}\left(\partial_if({\bm r})\partial_j+\partial_jf({\bm r})\partial_i\right),\nonumber\\
f({\bm r})\partial_i&\to&\frac{1}{2}\left(\partial_if({\bm r})+f({\bm r})\partial_i\right).\label{eq:recipe}
\end{eqnarray}
All the elements in the Hamiltonian matrix are a linear combination of
the elements of the form $E_1=f({\bm r})e_{ij}$, $E_2=f({\bm r})e_{ij}k_l$, 
$E_3=f({\bm r})x^\alpha y^\beta$, $E_4=f({\bm r})x^\alpha y^\beta k_i$ and $E_5=f({\bm r})x^\alpha y^\beta k_ik_j$, 
where $k_i$ ($i\in\{1,2,3\}$) is the differential operator $k_i=-i\frac{\partial}{\partial x_i}$,
$e_{ij}$ are the components of the strain tensor 
and $f({\bm r})$ is of the form 
\begin{equation}
f({\bm r})=f^{QD}\chi_{QD}({\bm r})+f^{M}(1-\chi_{QD}({\bm r})),
\end{equation}
where $f^{QD}$ is the value of a material parameter in the quantum dot and $f^{M}$ its value in the matrix,
$\chi_{QD}({\bm r})$ is the quantum dot characteristic function equal to 1 inside the dot
and 0 outside the dot. Their Fourier transforms (\ref{eq:fur}) are within the recipe (\ref{eq:recipe}) thus given by
\begin{equation}
\widetilde{E}_1({\bm q},{\bm k})=\frac{(2\pi)^3}{V}f^M\widetilde{e}_{ij}({\bm q}-{\bm k})-\frac{(2\pi)^6}{V^2}\Delta f
                      \sum_{\bm q'}\widetilde{\chi}_{QD}({\bm q}-{\bm k}-{\bm q'})\widetilde{e}_{ij}({\bm q'}),
\end{equation}
\begin{equation}
\widetilde{E}_2({\bm q},{\bm k})=\frac{1}{2}(k_l+q_l)\left[\frac{(2\pi)^3}{V}f^M\widetilde{e}_{ij}({\bm q}-{\bm k})-\frac{(2\pi)^6}{V^2}\Delta f
                      \sum_{\bm q'}\widetilde{\chi}_{QD}({\bm q}-{\bm k}-{\bm q'})\widetilde{e}_{ij}({\bm q'})\right],
\end{equation}
\begin{equation}
\widetilde{E}_3({\bm q},{\bm k})=f^MJ_{x^\alpha y^\beta}({\bm q}-{\bm k})-\frac{(2\pi)^3}{V}\Delta f\widetilde{\chi}^{x^\alpha y^\beta}_{QD}({\bm q}-{\bm k}),
\end{equation}
\begin{equation}
\widetilde{E}_4({\bm q},{\bm k})=\frac{1}{2}(k_i+q_i)\left[f^MJ_{x^\alpha y^\beta}({\bm q}-{\bm k})-\frac{(2\pi)^3}{V}\Delta f\widetilde{\chi}^{x^\alpha y^\beta}_{QD}({\bm q}-{\bm k})\right],
\end{equation}
\begin{equation}
\widetilde{E}_5({\bm q},{\bm k})=\frac{1}{2}(k_iq_j+q_ik_j)\left[f^MJ_{x^\alpha y^\beta}({\bm q}-{\bm k})-\frac{(2\pi)^3}{V}\Delta f\widetilde{\chi}^{x^\alpha y^\beta}_{QD}({\bm q}-{\bm k})\right],
\end{equation}
where $\Delta f=f^M-f^{QD}$, $\widetilde{e}_{ij}({\bm q})$
are Fourier transforms of the strain components given by
\begin{equation}
\widetilde{e}_{ij}({\bm q})=\frac{1}{(2\pi)^3}\int_V{\rm d}^3{\bm r}\exp\left(-i{\bm q}\cdot{\bm r}\right)e_{ij}({\bm r}),
\end{equation}
$\widetilde{\chi}_{QD}^{x^\alpha y^\beta}({\bm q})$ is Fourier transform of the quantum dot characteristic functions
\begin{equation}\label{eq:chi}
\widetilde{\chi}_{QD}^{x^\alpha y^\beta}({\bm q})=\frac{1}{(2\pi)^3}\int_{QD}{\rm d}^3{\bm r}\mrd x^\alpha y^\beta\exp\left(-i{\bm q}\cdot{\bm r}\right),
\end{equation}
where the integration goes only over the volume of the quantum dot and
\begin{equation}
J_{x^\alpha y^\beta}({\bm q}-{\bm k})=\delta_{k_z,q_z}\frac{1}{L_y}\int_{-L_y/2}^{L_y/2}e^{-i(q_y-k_y)y}y^\beta\mrd\textrm{d}y\frac{1}{L_x}\int_{-L_x/2}^{L_x/2}e^{-i(q_x-k_x)x}x^\alpha\mrd\textrm{d}x,
\end{equation}
where $\alpha$ and $\beta$ are non-negative integers from the set 
\begin{equation}
(\alpha,\beta)\in\{(0,0);(1,0);(2,0);(0,1),(1,1),(0,2)\}.
\end{equation}
The center of the pyramid base is taken as the origin of the coordinate system (Fig.~\ref{fig:1QD}).
The analytical formulae from which $\widetilde{e}_{ij}({\bm q})$ can be derived in a crystal
with cubic symmetry are given in Ref.~\onlinecite{jap86-297}. After integration, the characteristic 
functions can all be expressed as a linear combination of integrals
of the type
\begin{equation}
I_m(q)=\int_0^{b/2}x^me^{iqx}\mrd\textrm{d}x,
\end{equation}
where $m\in\{0,1,2,3\}$ and can be therefore evaluated analytically. All the
integrals $J_{x^\alpha y^\beta}({\bm q}-{\bm k})$ are evaluated analytically, as well.

In order to find the energy levels and the wave functions in the quantum dot, the eigenvalue problem (\ref{eq:eigen}) should
be solved. The direct application of this approach would lead to an eigenvalue problem of a matrix of size 
$N\times N$.  However, it is possible to exploit the symmetry 
of the model 
to block diagonalize
the corresponding matrix, as will be done in Sec.~\ref{sec:symmetry}.


\section{The symmetry of the model}\label{sec:symmetry}
As already discussed in the introduction, the symmetry group of the model is the double group $\overline{C}_4$. The generator of the
group is the total angular momentum $\hat{F}_z$ and therefore the representations of the
elements of the group are given by the operators $\hat{D}(R(\varphi))=\exp(-i\varphi\hat{F}_z)$, where 
$\varphi\in\{k\pi/2\}$ ($k\in\{0,1,\ldots,7\}$) and $R(\varphi)$ is a rotation by an angle $\varphi$.
In order to find how operators $\hat{D}(R(\varphi))$ act on the states (Eq.~\ref{eq:state}), it is enough to find how $\hat{D}(R(\pi/2))$
acts on the states since $\hat{D}(R(k\pi/2))=\hat{D}(R(\pi/2))^k$.
The total angular momentum $\hat{F}_z$ is a sum of the total angular momentum of the Bloch function $\hat{J}_z$
and the orbital angular momentum of the envelope function $\hat{L}_z$, i.e. $\hat{F}_z=\hat{J}_z+\hat{L}_z$\cite{prb55-4580}. 
Therefore action of the operator $\hat{D}(R(\pi/2))$ on state $|\Psi\rangle$ is composed of a  
rotation of the envelope functions in real space generated by its orbital angular momentum $\hat{L}_z$ and
a rotation of the Bloch function generated by its total angular momentum $\hat{J}_z$
\begin{equation}
\hat{D}(R(\pi/2))|\Psi\rangle=\sum_{i=1}^8 \left[\exp(-i\varphi\hat{L}_z)\psi_i({\bm r})\right]
					   \left[\exp(-i\varphi\hat{J}_z)|i\rangle\right].
\end{equation}
Since the basis of Bloch states $|i\rangle$ is the eigenbasis of $\hat{J}_z$ it follows that
\begin{equation}
\exp(-i\hat{J}_z\pi/2)|i\rangle=\exp(-iJ_z(i)\pi/2)|i\rangle,
\end{equation}
where $J_z(i)$ is the eigenvalue of the $z-$component of the total angular momentum of Bloch function $|i\rangle$
($J_z(1)=-1/2$, $J_z(2)=1/2$, $J_z(3)=-3/2$, $J_z(4)=-1/2$, $J_z(5)=1/2$, $J_z(6)=3/2$, $J_z(7)=-1/2$, $J_z(8)=1/2$). 
Thus the operator $\hat{D}(R(\pi/2))$ acts on the state $|\Psi\rangle$ as
\begin{equation}\label{eq:del}
\hat{D}(R(\pi/2))|\Psi\rangle=\sum_{i=1}^8 \psi_i(y,-x,z)\exp(-iJ_z(i)\pi/2)|i\rangle.
\end{equation}



By assuming the envelope functions as a linear combination of a finite number of plane waves, we have already 
reduced the otherwise infinite Hilbert space of the model
to the Hilbert space $\mathcal{H}$ of dimension $N$ formed by linear combination of plane waves multiplied by
the Bloch functions. The basis of the space $\mathcal{H}$ is given by
\begin{equation}\label{eq:bas}
|{\bm k},i\rangle=\exp(i{\bm k}\cdot{\bm r})|i\rangle,
\end{equation} 
where ${\bm k}=2\pi\left(\frac{m_x}{L_x},\frac{m_y}{L_y},\frac{m_z}{L_z}\right)$, $m_x\in\{-n_x,\ldots,n_x\}$,
 $m_y\in\{-n_y,\ldots,n_y\}$, $m_z\in\{-n_z,\ldots,n_z\}$ and $i\in\{1,2,\ldots,8\}$.

We shall first represent the operator $\hat{D}(R(\pi/2))$ in the plane wave basis of the space $\mathcal{H}$. We
thus need to know how $\hat{D}(R(\pi/2))$ acts on the basis vectors. Using (\ref{eq:del}) and (\ref{eq:bas}) we
find that
\begin{equation}
\hat{D}(R(\pi/2))|(k_x,k_y,k_z),i\rangle=\exp(-iJ_z(i)\pi/2)|(-k_y,k_x,k_z),i\rangle.
\end{equation}
One can note that for $(k_x,k_y)=(0,0)$ the acting of $\hat{D}(R(\pi/2))$ on the basis vector is just a phase
shift and the orbit of acting of the group elements is just an one-dimensional space (we shall denote it
as $\mathcal{H}_{(0,0,k_z),i}$), while for 
$(k_x,k_y)\ne(0,0)$ the orbit is a four-dimensional space (that we shall denote 
as $\mathcal{H}_{(k_x,k_y,k_z),i}$, where $k_x>0$ and $k_y\ge 0$ to avoid multiple counting
of the same space) with the basis 
\begin{eqnarray}\label{eq:nbas}
|b1\rangle&=&|(k_x,k_y,k_z),i\rangle,\nonumber\\ 
|b2\rangle&=&|(-k_y,k_x,k_z),i\rangle,\nonumber \\ 
|b3\rangle&=&|(-k_x,-k_y,k_z),i\rangle,\nonumber \\ 
|b4\rangle&=&|(k_y,-k_x,k_z),i\rangle. 
\end{eqnarray}
In the space $\mathcal{H}_{(0,0,k_z),i}$ the representation $\hat{D}$ reduces to an one-dimensional
representation defined by 
\begin{equation}\label{eq:repr1}
\hat{D}_{(0,0,k_z),i}(R(\pi/2))=\exp(-iJ_z(i)\pi/2), 
\end{equation}
while in the space $\mathcal{H}_{(k_x,k_y,k_z),i}$ it reduces to a four-dimensional representation which
is given in the basis from Eq.~\ref{eq:nbas} by 
\begin{equation}\label{eq:repr}
\hat{D}_{(k_x,k_y,k_z),i}(R(\pi/2))=\exp(-iJ_z(i)\pi/2)\left[\begin{array}{cccc}
       0     &    0     &    0     &     1    \\
       1     &    0     &    0     &     0    \\
       0     &    1     &    0     &     0    \\
       0     &    0     &    1     &     0    \\	     	     
\end{array}\right].
\end{equation}
Since the spaces $\mathcal{H}_{(k_x,k_y,k_z),i}$ and $\mathcal{H}_{(0,0,k_z),i}$ are invariant 
for the representation $\hat{D}$, it is given by an orthogonal sum  
\begin{equation}\label{eq:D0}
\hat{D}=\bigoplus_{k_x,k_y,k_z,i}\hat{D}_{(k_x,k_y,k_z),i}+\bigoplus_{k_z,i}\hat{D}_{(0,0,k_z),i}.
\end{equation}
From Eqs. \ref{eq:repr1} and \ref{eq:repr} one finds that the characters of the representation 
of the group elements are given by
\begin{equation}\label{eq:D1}
\chi\left(\hat{D}_{(k_x,k_y,k_z),i}(R(k\pi/2))\right)=
\left\{\begin{array}{cc}
  4   &  k=0    \\
 -4   &  k=4   \\
  0   &  k\in\{1,2,3,5,6,7\}
\end{array}\right.
\end{equation}
and
\begin{equation}\label{eq:D2}
\chi\left(\hat{D}_{(0,0,k_z),i}(R(k\pi/2))\right)=\exp(-iJ_z(i)k\pi/2).
\end{equation}

The characters of the irreducible representations of the double
group $\overline{C}_4$ are given by $\chi\left(A_{m_f}(R(k\pi/2))\right)=\exp(ikm_f\pi/2)$,
where $m_f\in\{-3/2,-1,-1/2,0,1/2,1,3/2,2\}$ and $k\in\{0,1,\ldots,7\}$. One finds from (\ref{eq:D1})
that
\begin{equation}\label{eq:DD1}
\hat{D}_{(k_x,k_y,k_z),i}=A_{1/2}+A_{-1/2}+A_{3/2}+A_{-3/2}
\end{equation}
and from (\ref{eq:D2}) obviously
\begin{equation}\label{eq:DD2}
\hat{D}_{(0,0,k_z),i}=A_{-J_z(i)}.
\end{equation}
Using (\ref{eq:DD1}) and (\ref{eq:DD2}), it follows from (\ref{eq:D0}) that
\begin{equation}
\hat{D}=N_1A_{1/2}+N_1A_{-1/2}+N_2A_{3/2}+N_2A_{-3/2},
\end{equation}
where
\begin{equation}
N_1=8n_x(n_y+1)(2n_z+1)+3(2n_z+1)
\end{equation}
and
\begin{equation}
N_2=8n_x(n_y+1)(2n_z+1)+2n_z+1.
\end{equation}

Projection operators \cite{Elliott} were then used to find the symmetry adapted
basis. 
The projection operators are given by
\begin{equation}\label{eq:projector}
\hat{P}_{A_{m_f}}((k_x,k_y,k_z),i)=\frac{1}{8}\sum_{k=0}^{7}\chi\left(A_{m_f}(R(k\pi/2))\right)^* \hat{D}_{(k_x,k_y,k_z),i}(R(k\pi/2)),
\end{equation}
while 
\begin{equation}
\hat{P}_{A_{m_f}}((0,0,k_z),i)=1
\end{equation}
and they project arbitrary states in space $\mathcal{H}$ to the elements of the symmetry adapted basis.
The explicit forms of the projection operators are derived from (\ref{eq:projector}) and (\ref{eq:repr})
and in the basis (\ref{eq:nbas}) are equal to
\begin{eqnarray}
\hat{P}_{A_{-3/2}}(1)=\hat{P}_{A_{-3/2}}(4)&=&\hat{P}_{A_{-3/2}}(7)=\hat{P}_{A_{-1/2}}(3)=\nonumber\\
=\hat{P}_{A_{1/2}}(6)&=&\hat{P}_{A_{3/2}}(2)=\hat{P}_{A_{3/2}}(5)=\hat{P}_{A_{3/2}}(8)=M_1,\nonumber\\
\hat{P}_{A_{-3/2}}(2)=\hat{P}_{A_{-3/2}}(5)&=&\hat{P}_{A_{-3/2}}(8)=\hat{P}_{A_{-1/2}}(1)=\nonumber\\
=\hat{P}_{A_{-1/2}}(4)&=&\hat{P}_{A_{-1/2}}(7)=\hat{P}_{A_{1/2}}(3)=\hat{P}_{A_{3/2}}(6)=M_2,\nonumber\\
\hat{P}_{A_{-3/2}}(6)=\hat{P}_{A_{-1/2}}(2)&=&\hat{P}_{A_{-1/2}}(5)=\hat{P}_{A_{-1/2}}(8)=\nonumber\\
=\hat{P}_{A_{1/2}}(1)&=&\hat{P}_{A_{1/2}}(4)=\hat{P}_{A_{1/2}}(7)=\hat{P}_{A_{3/2}}(3)=M_3,\nonumber\\
\hat{P}_{A_{-3/2}}(3)=\hat{P}_{A_{-1/2}}(6)&=&\hat{P}_{A_{1/2}}(2)=\hat{P}_{A_{1/2}}(5)=\nonumber
\\=\hat{P}_{A_{1/2}}(8)&=&\hat{P}_{A_{3/2}}(1)=\hat{P}_{A_{3/2}}(4)=\hat{P}_{A_{3/2}}(7)=M_4,
\end{eqnarray}
where $(k_x,k_y,k_z)$ was omitted in all brackets for brevity and where 
the matrices $M_1$, $M_2$, $M_3$ and $M_4$ are given by
\begin{eqnarray*}
M_1=\frac{1}{4}\left[
\begin{array}{cccc}
1 & -1 & 1 & -1 \\
-1 & 1 & -1 & 1 \\
1 & -1 & 1 & -1 \\
-1 & 1 & -1 & 1 
\end{array}
\right], &
M_2=\frac{1}{4}\left[
\begin{array}{cccc}
1 & -i & -1 & i \\
i & 1 & -i & -1 \\
-1 & i & 1 & -i \\
-i & -1 & i & 1 
\end{array}
\right], 
\end{eqnarray*}
\begin{eqnarray}
M_3=\frac{1}{4}\left[
\begin{array}{cccc}
1\mrd & 1\mrd & 1\mrd & 1\mrd \\
1\mrd & 1\mrd & 1\mrd & 1\mrd \\
1\mrd & 1\mrd & 1\mrd & 1\mrd \\
1\mrd & 1\mrd & 1\mrd & 1\mrd 
\end{array}
\right], &
M_4=\frac{1}{4}\left[
\begin{array}{cccc}
1 & i & -1 & -i \\
-i & 1 & i & -1 \\
-1 & -i & 1 & i \\
i & -1 & -i & 1 
\end{array}
\right].
\end{eqnarray}

The elements of the symmetry adapted basis are finally given as:
\begin{eqnarray}
|A_{1/2},(0,0,k_z),i\rangle=&|(0,0,k_z),i\rangle & i\in\{1,4,7\} \nonumber\\
|A_{1/2},(k_x,k_y,k_z),i\rangle=&\frac{1}{2}\left(|b1\rangle+|b2\rangle+|b3\rangle+|b4\rangle \right) & i\in\{1,4,7\} \nonumber\\
|A_{1/2},(k_x,k_y,k_z),i\rangle=&\frac{1}{2}\left(|b1\rangle-i|b2\rangle-|b3\rangle+i|b4\rangle\right) & i\in\{2,5,8\} \nonumber\\
|A_{1/2},(k_x,k_y,k_z),i\rangle=&\frac{1}{2}\left(|b1\rangle-|b2\rangle+|b3\rangle-|b4\rangle\right) & i=6 \nonumber\\
|A_{1/2},(k_x,k_y,k_z),i\rangle=&\frac{1}{2}\left(|b1\rangle+i|b2\rangle-|b3\rangle-i|b4\rangle\right) & i=3,
\end{eqnarray}
\begin{eqnarray}
|A_{-1/2},(0,0,k_z),i\rangle=&|(0,0,k_z),i\rangle & i\in\{2,5,8\} \nonumber\\
|A_{-1/2},(k_x,k_y,k_z),i\rangle=&\frac{1}{2}\left(|b1\rangle+|b2\rangle+|b3\rangle+|b4\rangle\right) & i\in\{2,5,8\} \nonumber\\
|A_{-1/2},(k_x,k_y,k_z),i\rangle=&\frac{1}{2}\left(|b1\rangle-i|b2\rangle-|b3\rangle+i|b4\rangle\right) &  i=6 \nonumber\\
|A_{-1/2},(k_x,k_y,k_z),i\rangle=&\frac{1}{2}\left(|b1\rangle-|b2\rangle+|b3\rangle-|b4\rangle\right) & i=3 \nonumber\\
|A_{-1/2},(k_x,k_y,k_z),i\rangle=&\frac{1}{2}\left(|b1\rangle+i|b2\rangle-|b3\rangle-i|b4\rangle\right) & i\in\{1,4,7\},
\end{eqnarray}
\begin{eqnarray}
|A_{-3/2},(0,0,k_z),i\rangle=&|(0,0,k_z),i\rangle & i=6 \nonumber\\
|A_{-3/2},(k_x,k_y,k_z),i\rangle=&\frac{1}{2}\left(|b1\rangle+|b2\rangle+|b3\rangle+|b4\rangle\right) &  i=6\nonumber\\
|A_{-3/2},(k_x,k_y,k_z),i\rangle=&\frac{1}{2}\left(|b1\rangle-i|b2\rangle-|b3\rangle+i|b4\rangle\right) & i=3 \nonumber\\
|A_{-3/2},(k_x,k_y,k_z),i\rangle=&\frac{1}{2}\left(|b1\rangle-|b2\rangle+|b3\rangle-|b4\rangle\right) & i\in\{1,4,7\}\nonumber \\
|A_{-3/2},(k_x,k_y,k_z),i\rangle=&\frac{1}{2}\left(|b1\rangle+i|b2\rangle-|b3\rangle-i|b4\rangle\right) & i\in\{2,5,8\} ,
\end{eqnarray}
\begin{eqnarray}
|A_{3/2},(0,0,k_z),i\rangle=&|(0,0,k_z),i\rangle & i=3 \nonumber\\
|A_{3/2},(k_x,k_y,k_z),i\rangle=&\frac{1}{2}\left(|b1\rangle+|b2\rangle+|b3\rangle+|b4\rangle\right) &  i=3 \nonumber\\
|A_{3/2},(k_x,k_y,k_z),i\rangle=&\frac{1}{2}\left(|b1\rangle-i|b2\rangle-|b3\rangle+i|b4\rangle\right) & i\in\{1,4,7\}\nonumber \\
|A_{3/2},(k_x,k_y,k_z),i\rangle=&\frac{1}{2}\left(|b1\rangle-|b2\rangle+|b3\rangle-|b4\rangle\right) & i\in\{2,5,8\} \nonumber \\
|A_{3/2},(k_x,k_y,k_z),i\rangle=&\frac{1}{2}\left(|b1\rangle+i|b2\rangle-|b3\rangle-i|b4\rangle\right) & i=6. 
\end{eqnarray}
 
The Hamiltonian matrix elements between basis elements having different symmetry are equal to zero
implying that in this basis the Hamiltonian matrix is block diagonal with four blocks of sizes 
$N_1\times N_1$, $N_1\times N_1$, $N_2\times N_2$ and $N_2\times N_2$, respectively. The time necessary
to diagonalize the matrix of size $N\times N$ scales approximately as $N^3$. Therefore the block diagonalization
obtained reduces the computational time approximately by a factor of 16. Since all the basis
vectors of the symmetry adapted basis are linear combinations of one or four vectors of the plane wave basis, 
it follows that Hamiltonian matrix elements in the symmetry adapted basis can be expressed as linear combinations of 
one, four or sixteen Hamiltonian matrix elements in the plane wave basis. Therefore all the elements of the 
four blocks are given by analytical formulae that can be easily derived from the analytical formulae for the 
matrix elements in the plane wave basis given in Sec.~\ref{sec:pwm}.

It can be proved 
by considering the two dimensional irreducible representations of the more general double group $\overline{C}_{4v}$
\cite{Elliott} that states with the same absolute value of $m_f$ are degenerate in pairs (the Kramer's degeneracy).
This degeneracy is lifted in the presence of external axial magnetic field $B$ when the symmetry reduces
from $\overline{C}_{4v}$ to $\overline{C}_{4}$ and the time reversal symmetry relation $E_{m_f}(B)=E_{-m_f}(-B)$ holds then.

It has been pointed out \cite{prb58-9955, prb52-11969} that piezoelectric effects in single dots of realistic sizes
are small, changing the eigen-energies of the system by less than $1\mrd{\rm meV}$, and can be neglected. It has also been 
shown that in a vertically stacked double quantum dot the influence of a piezoelectric field is more important since the piezoelectric potential
generated by the two dots adds up in the regions above and below the dots, while it is almost cancelled out in the region between the dots.
Consequently, it is expected that in a periodic array of vertically stacked quantum dots considered in Sec.~\ref{sec:results} the piezoelectric potential in the region between the dots would be almost cancelled out and that the influence of piezoelectric effect on eigen-energies would be small. 
Therefore the small piezoelectric potential that breaks the symmetry of the system from $\overline{C}_4$ to $\overline{C}_2$ can be treated as
a perturbation. It belongs to the $A_2$ representation of $\overline{C}_4$ group and therefore
only the matrix elements between the states with $\Delta m_f=2$ are non zero. Consequently, the piezoelectric potential doesn't change the 
energies in the first order of the perturbation theory and second order perturbation theory is needed to take the piezoelectric effect into account.

We introduce the following notation for the electron states $ne_{m_f}$, where $n$ is a positive integer labeling the states 
with given $m_f$ in increasing order of their energy. Since the quantum number $m_f$ originates from the irreducible
representations of the double group $\overline{C}_4$ whose elements are generated by the total angular momentum, its physical 
interpretation is that it represents the total quasi-angular momentum. The hole states 
will be labeled by $nh_{m_f}$, with the same meaning of the symbol as for the electron case, except that $n$ labels the states
in decreasing order of their energy, as is natural for holes.

The Hamiltonian of the interaction with the electromagnetic field is obtained by replacing ${\bm k}$ with ${\bm k+\frac{e}{\hbar}{\bm A}}$
in the kinetic part of the Hamiltonian. In the dipole approximation ${\bm A}$ can be considered constant in space
and furthermore all the terms quadratic in ${\bm A}$ are neglected. After calculating the matrix elements between
states with a well defined symmetry, we obtain the following selection rules. If the light is $z-$polarized then 
$\Delta m_f=0$, while if the light is $\sigma^\pm$ circularly polarized then $\Delta m_f=\pm 1$ (where by 
definition $3/2+1=-3/2$ and $-3/2-1=3/2$).

\section {Results}\label{sec:results}
\begin{figure}[htbp]
\vspace{1cm}
\includegraphics[width=0.5\textwidth]{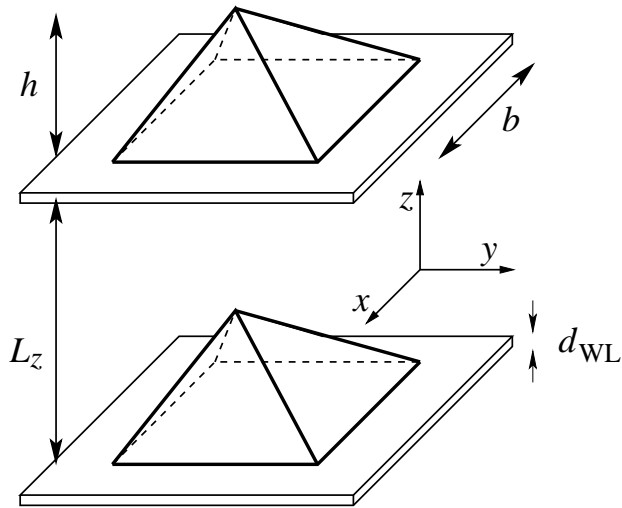}
\caption[]{Geometry of a periodic array of pyramidal quantum dots. The width of the pyramid base is $b$, the height $h$, 
the wetting layer width $d_{WL}$, the period of the structure is $L_z$.}
\label{fig:geometry}
\end{figure}

The method presented was applied to the calculation of the electronic structure of periodic array of
vertically stacked pyramidal self-assembled quantum dots (Fig.~\ref{fig:geometry}). The dimensions of the dots in an array were
taken to be equal to those estimated for the structure reported in Ref. \onlinecite{prl76-952} - the base width $b=18\mrd\textrm{nm}$, 
the height $h=4\mrd\textrm{nm}$, the wetting layer width $d_{WL}=1.7\mrd\textrm{ML}$, while the period 
of the structure in $z-$direction $L_z$ was varied in the interval from $L_z=h+d_{WL}$ where the dots lie
on top of one another to $L_z=16\mrd\textrm{nm}$. The dimensions of the embedding box $L_x=L_y=2b$ were 
taken.
The material parameters were taken from Ref.~\onlinecite{jap89-5815}.
According to Bloch's theorem, the $k-$th component of the state
spinor is given by 
\begin{equation}
\Psi_k({\bm r})=\exp(iK_zz)\psi_k({\bm r}),
\end{equation}
where $k\in\{1,2,\ldots,8\}$ and $\psi_k({\bm r})$ is periodic in the $z-$direction with the period $L_z$. 
Therefore, the matrix elements in the four blocks of the Hamiltonian matrix 
for calculating $E(K_z)$ are given by linear combinations of the elements obtained by the same formulae
from Sec.~\ref{sec:pwm} except that $k_z$ and $q_z$ should be replaced by $k_z+K_z$ and $q_z+K_z$, 
respectively. Since the relation $E(K_z)=E(-K_z)$ holds, only the states with $K_z\ge 0$ will be considered.
The InAs unstrained conduction band edge is taken as the energy reference level. 

The small piezoelectric effect was assumed to be negligible in the calculation. In order to check this assumption
it was included in the framework of second order perturbation theory (Sec.~\ref{sec:symmetry}). Its 
influence on the state energies was of the order of $1\mrd{\rm meV}$ and less, confirming our assumption. 

Two main factors influence the electronic structure of the periodic array of quantum dots~-~strain distribution and
quantum mechanical coupling. 

The influence of quantum mechanical coupling is intuitively clear - as the distance between 
the dots increases the coupling is weaker implying smaller miniband widths. Due to their large effective mass, heavy-holes 
are the least influenced by coupling and the minibands of dominantly heavy-hole like states are narrow, while the minibands 
of electron and light-hole states are much wider.

On the other hand, the strain distribution is complex and in principle all six components of the strain tensor 
influence the electronic structure. Still, the most important are hydrostatic strain $e_h=e_{11}+e_{22}+e_{33}$ 
that determines the position of the electron and hole levels and biaxial strain $e_b=e_{33}-\frac{1}{2}\left(e_{11}+e_{22}\right)$ 
whose main influence is on splitting of the light and heavy-hole states. The bigger the value of hydrostatic strain, 
the lower the conduction band states are in energy and the higher the valence band states are in energy. When the biaxial strain 
is negative, the light-holes tend to have higher energy than heavy-holes, while when it is positive the situation is opposite.
Having the importance of those two components of strain in mind, 
we have investigated first how they 
change when the distance between the dots in an array is varied. 
We have found that as the distance between the dots increases, the hydrostatic strain in the dots decreases. On the other hand,
for small values of the period, the biaxial strain is negative, while for larger values it changes sign and
increases further.

\subsection{Energy levels in the conduction band}
\begin{figure}[htbp]
\vspace{1cm}
\includegraphics[width=0.7\textwidth]{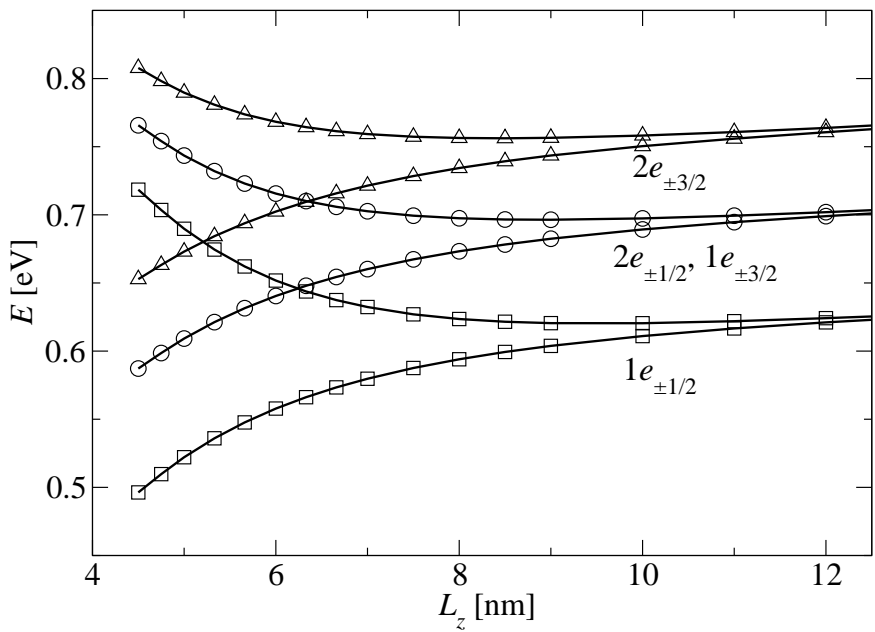}
\caption{The conduction miniband minima and maxima dependence on the period of the structure $L_z$. 
The $1e_{\pm 1/2}$ miniband is represented by squares, the $1e_{\pm 3/2}$ and $2e_{\pm 1/2}$ miniband by circles, the 
$2e_{\pm 3/2}$ miniband by triangles.}
\label{fig:el18-4}
\end{figure}

The dependence of the miniband minima and maxima on the period of the structure is given in Fig.~\ref{fig:el18-4}.
This behavior is expected. When the dots are close, they are strongly coupled and the minibands are wide,
while as $L_z$ increases the coupling is weaker and the energy spectrum becomes discrete. For large values of 
the period when the miniband width practically vanishes, we still see a rising trend in energy. This rise
is caused by a still decreasing value of hydrostatic strain.
This leads us to the conclusion that the range of strain effects is larger than the range of quantum mechanical
coupling.

The ground miniband
has $|m_f|=1/2$ symmetry, while the first and second excited miniband having different symmetries $|m_f|=1/2$
and $|m_f|=3/2$ are nearly degenerate. Their difference in energy is less than $1\mrd\textrm{meV}$, too small 
to be seen on the graph. 
A comment should be given about the near degeneracy of those two states. It has been
practice in the literature to say that these two states are exactly degenerate in the absence of a piezoelectric 
effect and that the piezoelectric effect breaks the degeneracy of these states. This is indeed
true if the carrier energy spectrum in the quantum dot is modeled by simple one band Schr\"{o}dinger equation. 
The symmetry group is then $C_{4v}$, consisting of transformations in real space generated by the orbital angular 
momentum $\hat{L}_z$ and the first and second excited state transform according to the same two dimensional
irreducible representation of $C_{4v}$ implying their degeneracy. However, when the 8-band $k\cdot p$ model,
which is inherently spin-dependent, is used, the Hamiltonian no longer commutes with the rotations generated by
orbital angular momentum, but the total angular momentum. The symmetry group is the double $\overline{C}_4$ group that has only the one dimensional irreducible representations and
there is no a priori reason for the states with different absolute values of $m_f$ to be degenerate.
  
In order to explain the near degeneracy of the states with different symmetry, we have further investigated
the 8-band $k\cdot p$ Hamiltonian and checked that it would still commute with the transformations in real space
generated by orbital angular momentum if the valence band spin-orbit splitting would be set to zero. 
Since the influence of valence band spin-orbit splitting on the levels in the conduction band is not
substantial, the degeneracy of the two states is small. We thus conclude that the origin of splitting of 
the first and the second excited state is not just the piezoelectric effect but also the valence band spin-orbit 
splitting.


All the minibands shown in Fig.~\ref{fig:el18-4} exhibit minima at $K_z=0$ and maxima at $K_z=\pi/L_z$
for all values of the period $L_z$. For small values of the period $L_z$ there is an energy overlap between
different minibands, while the minibands are completely separated for larger values of $L_z$.
There is no crossing between states of different symmetry. 

\subsection {Energy levels in the valence band}
\begin{figure}[htbp]
\vspace{1cm}
\includegraphics[width=0.7\textwidth]{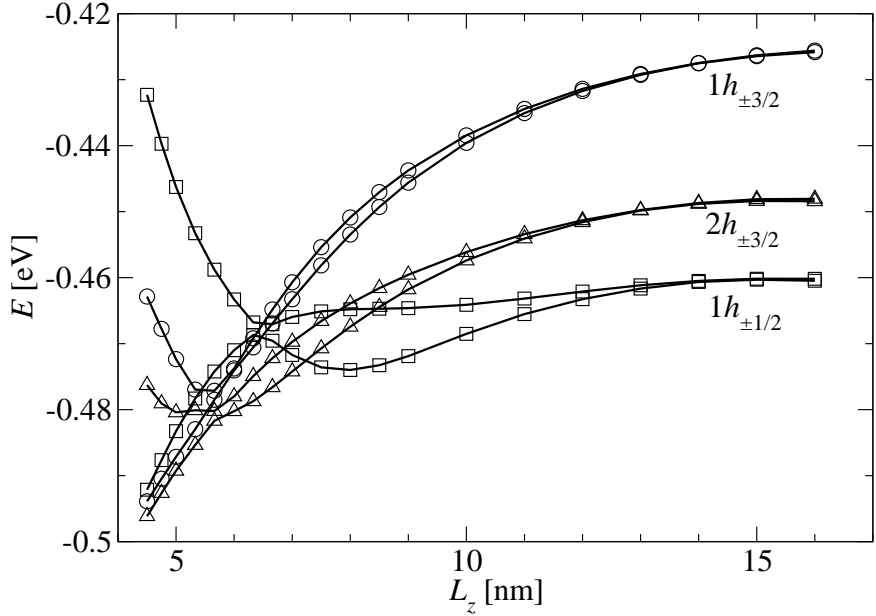}
\caption{The miniband minima and maxima dependence on the period of the structure $L_z$. The  
$1h_{\pm 1/2}$ miniband is represented by squares, the $1h_{\pm 3/2}$ miniband by circles and
the $2h_{\pm 3/2}$ miniband by triangles.}
\label{fig:ho18-4}
\end{figure}

The miniband minima and maxima dependence on the period of the structure are given in Fig.~\ref{fig:ho18-4}
for the three highest minibands in the valence band. Due to the combined effects of strain, mixing of 
light and heavy-holes and quantum mechanical coupling between the dots, the hole minibands exhibit a more
complex structure than the electron minibands. 

\begin{figure}[htbp]
\vspace{1cm}
\includegraphics[width=\textwidth]{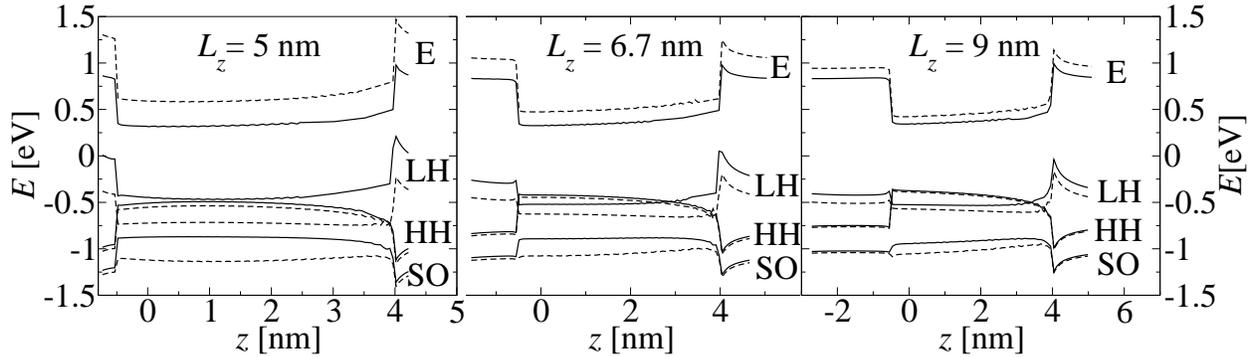}
\caption{ Effective potentials at three different values of the period $L_z$
 at $K_z=0$ (full lines) and $K_z=\pi/L_z$ (dashed lines) for electrons (E),
 light-holes (LH), heavy-holes (HH) and spin orbit split holes (SO).}
\label{fig:veff}
\end{figure}

In order to explain such behavior we note first that the effective potential felt by carriers depends
on $K_z$. We define the effective potential as the value obtained by diagonalizing the Hamiltonian with
$k_x=k_y=0$ and $k_z=K_z$. The $z$-dependence of electron, light, heavy and spin-orbit split hole 
$K_z=0$ and $K_z=\pi/L_z$ effective potentials at the pyramid axis for a few different values of $L_z$ is shown in 
Fig.~\ref{fig:veff}.

Since the effective mass of the light-holes is small, the light-hole effective 
potential is substantially different for $K_z=0$ and $K_z=\pi/L_z$, while in the case of heavy-holes
that difference is much smaller. As seen from Fig.~\ref{fig:veff}, as the period of 
the structure increases, the effective potential felt by $K_z=0$ light-holes decreases, while quite 
oppositely the effective potential felt by heavy-holes increases. Both of these trends are an expected 
consequence of the increase in the value of the biaxial strain.
Consequently, in the range of low values of $L_z$ the hole states with $K_z=0$
are dominantly of the light-hole type, while the states with $K_z=\pi/L_z$ are dominantly heavy-hole like.
The states with $K_z=\pi/L_z$ remain of heavy-hole type across the whole investigated interval of $L_z$ and 
their energy therefore increases with increasing $L_z$. The energy of the light-hole $K_z=0$ states decreases
with increasing $L_z$ and at the same time their heavy-hole content increases. As a consequence of two 
different energy trends for $K_z=0$ and $K_z=\pi/L_z$ the miniband width decreases for all the states
until a certain point where the energy of the $K_z=0$ state becomes less than the energy of the $K_z=\pi/L_z$ 
state. This point, where the inversion of the sign of the miniband effective mass occurs is different for
different states, for the $1h_{\pm 1/2}$ state it occurs around $L_z=6.5\mrd\textrm{nm}$, while for the
$1h_{\pm 3/2}$ and $2h_{\pm 3/2}$ states it occurs around $L_z=5.8\mrd\textrm{nm}$. 
The light-hole content of the $K_z=0$ states decreases with increasing $L_z$ and eventually they become 
dominantly heavy-hole like. This light to heavy-hole transition occurs at $L_z=7.5\mrd\textrm{nm}$ for 
the $1h_{\pm 1/2}$ state, at $L_z=5.5\mrd\textrm{nm}$ for the $1h_{\pm 3/2}$ state and 
$L_z=5.0\mrd\textrm{nm}$ for the $2h_{\pm 3/2}$ state. Since the energy of the light-hole states 
decreases with increasing $L_z$ and the energy of the heavy-hole states increases, the position of the light to heavy-
hole transition corresponds approximately to the position of the energy minima of $K_z=0$ states, as can 
be verified from Fig.~\ref{fig:ho18-4}. We further observe that the ground hole state for $L_z\le 6.5\mrd\textrm{nm}$
is $1h_{\pm 1/2}$ having $|m_f|=1/2$ symmetry, while for $L_z>6.5\mrd\textrm{nm}$ it is $1h_{\pm 3/2}$ having $|m_f|=3/2$ 
symmetry. Therefore at the
critical point $L_z=6.5\mrd\textrm{nm}$, we observe an interesting effect of a simultaneous change of ground 
hole state symmetry, a change of the sign of the effective mass and a change of the ground state type from light 
to heavy-hole like.

The spin orbit split band certainly influences the exact positions of the energy levels,
however, being far in energy from the light and heavy-hole bands it doesn't influence the overall behavior 
described in the previous paragraph. This is verified by the fact that the spin orbit band content of the hole
states is typically of the order of 5\%.

As far as the spatial localization of the wave functions is concerned, one would expect from the effective potential profiles given in Fig.~\ref{fig:veff} that dominantly light-hole states would be confined outside the dots and dominantly heavy-hole states inside the dots. However, the states are of light-hole type only when the dots are very close to each other and the effective potential well is then too narrow to confine the hole. Therefore, the light-hole like states are spread both inside and outside the dots. When the distance between the dots increases and light to heavy-hole transitions take place, the hole state becomes localized inside the dots.

\subsection {Influence of external axial magnetic field}

\begin{figure}[htbp]
\vspace{1cm}
\includegraphics[width=0.7\textwidth]{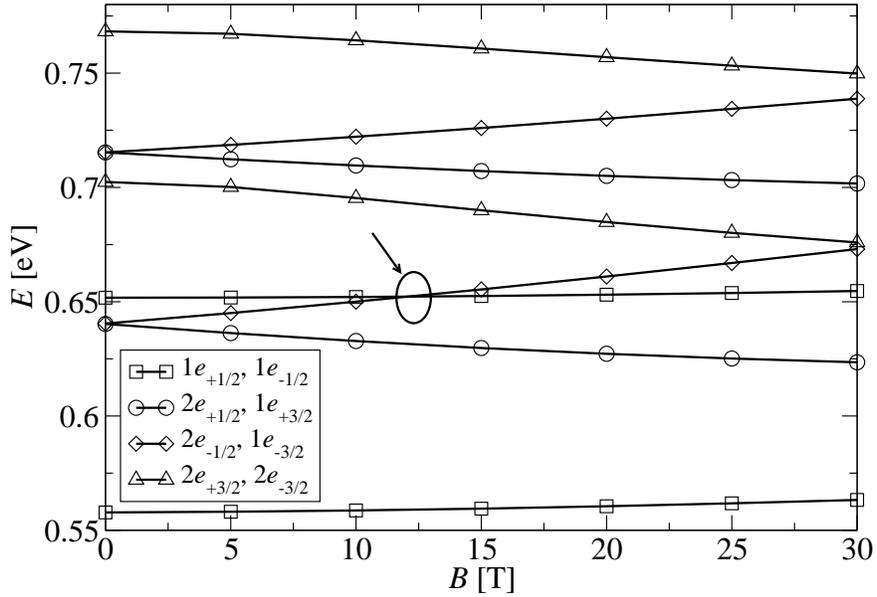}
\caption{Magnetic field dependence of miniband minima and maxima for $1e_{-3/2}$, $1e_{-1/2}$, $1e_{+1/2}$, $1e_{+3/2}$,
$2e_{-3/2}$, $2e_{-1/2}$, $2e_{+1/2}$ and $2e_{+3/2}$ states. The position where different minibands separate is marked.}
\label{fig:eleb}
\end{figure}

\begin{figure}[htbp]
\vspace{1cm}
\includegraphics[width=0.7\textwidth]{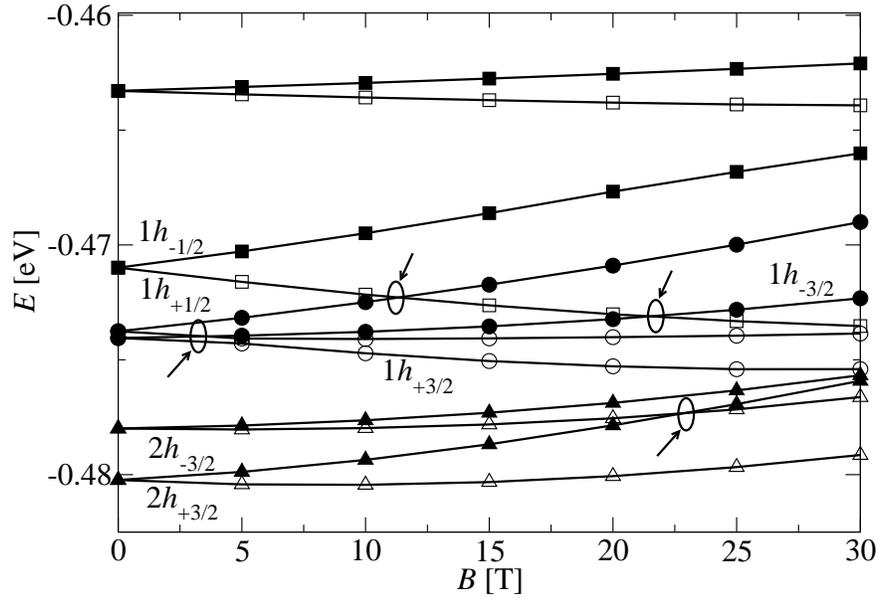}
\caption{Magnetic field dependence of miniband minima and maxima for $1h_{-3/2}$ (full circles), $1h_{+3/2}$ (empty circles), 
$1h_{-1/2}$ (full squares), $1h_{+1/2}$ (empty squares), $2h_{-3/2}$ (full triangles) and $2e_{+3/2}$ (empty triangles) states.
The positions where different minibands separate or concatenate are marked.}
\label{fig:holb}
\end{figure}

The magnetic field dependences of the miniband minima and maxima of the conduction and valence band states 
for the structure with the period $L_z=6\mrd\textrm{nm}$ are shown in Figs. \ref{fig:eleb} and \ref{fig:holb}, respectively.
As already mentioned, Kramer's degeneracy is broken in magnetic field. The relation $E_{m_f}(B)=E_{-m_f}(-B)$ holds, 
thus only the $B\ge 0$ part of the dependence is shown on the graphs. 

The magnetic field splitting between $1e_{+3/2}$ and $1e_{-3/2}$ states, as well as between $2e_{-1/2}$ and 
$2e_{+1/2}$ states is significant because the mesoscopic angular momentum~\cite{prb55-9275} of those states 
is different from zero. However, the splitting between $1e_{-1/2}$ and $1e_{+1/2}$ states and between $2e_{+3/2}$
and $2e_{-3/2}$ is much smaller, too small to be seen on the graph (of the order of few meV). There is no crossing 
between the states of different symmetry. We also note that the $1e_{+1/2}$ and $1e_{-1/2}$ minibands overlap with
$2e_{-1/2}$ and $1e_{-3/2}$ minibands for $B=0$ but as the magnetic field is increased this overlap vanishes (for $B\gtrsim 12\mrd\textrm{T}$). 
The energy separation of the minibands has an important affect on the dynamical characteristics of the structure since
it suppresses all the one particle energy conserving scattering mechanisms between those minibands (like ionized impurity scattering)
and with further separation even suppresses the mechanisms with energy exchange (like acoustic phonon and longitudinal optical
phonon scattering).

The splitting between the hole states is also of the order of a few meV, however since the energy difference between different 
states is also small, this splitting is enough to cause crossings between states of different symmetry. 
We further find from Fig.~\ref{fig:holb} that the minibands
$1h_{-3/2}$ and $1h_{+3/2}$ that are degenerate at $B=0$ become completely separated already at $B\gtrsim 3\mrd\textrm{T}$ and 
the same effect for $2h_{-3/2}$ and $2h_{+3/2}$ occurs at $B\gtrsim 23\mrd\textrm{T}$. Although the magnetic field splitting is
the most pronounced for $1h_{-1/2}$ and $1h_{+1/2}$ states, the effect of the separation of those minibands occurs at magnetic 
fields larger than $30\mrd\textrm{T}$, which is a consequence of larger miniband width than in the previous cases. Apart from 
separation of the minibands, the magnetic field can also concatenate otherwise nonoverlapping minibands. 
It is seen in Fig.~\ref{fig:holb} that $1h_{+1/2}$ and $1h_{-3/2}$ start to overlap at $B\sim 11\mrd\textrm{T}$ and that for
$B\gtrsim 22\mrd\textrm{T}$ the range of energies of $1h_{-3/2}$ becomes a subset of the range of energies of $1h_{+1/2}$ miniband.

\section{Conclusion}
In conclusion, we have developed a symmetry-based method for calculation of electronic states in pyramidal
InAs/GaAs quantum dots. The corresponding Hamiltonian matrix obtained by the plane wave method was block diagonalized
into four matrices of approximately equal size, which enabled significantly faster calculation of energy levels
within the plane wave method. The symmetry considerations not only enabled more efficient calculation of the 
electronic structure but also give more insight about the physics of the model by introducing the quantum number 
of total quasi-angular momentum and giving the selection rules for interaction with electromagnetic field.
The method developed was applied to calculate the electronic structure of a periodic array of vertically stacked pyramidal 
self-assembled quantum dots. It was found that as the distance between the dots is increased, at a certain critical point
the ground hole state simultaneously changes symmetry from $|m_f|=1/2$ to $|m_f|=3/2$ and type from light to heavy-hole.
The influence of magnetic field on the energy levels is in general less pronounced than the influence of quantum mechanical coupling and strain
but nevertheless it can be used for fine tuning of the properties of the structure since its increase or decrease is able
to separate energy overlapping minibands. 


\end{document}